\documentclass[prl,twocolumn,showpacs,superscriptaddress]{revtex4}
\usepackage{amsmath}
\usepackage{amsfonts}
\usepackage{amssymb}
\usepackage{graphicx}
\begin{document}

\title{Two-photon exchange in elastic electron-proton scattering : 
QCD factorization  approach}
\author{Nikolai Kivel}
\address{Institute f\"ur Theoretische Physik II,  Ruhr-Universit\"at Bochum, D-44780 Bochum, Germany}
\address{Petersburg Nuclear Physics Institute, 188350  Gatchina,  Russia}
\author{Marc Vanderhaeghen}
\affiliation{Institut f\"ur Kernphysik, Johannes Gutenberg-Universit\"at,
D-55099 Mainz, Germany}
\date{\today}
\begin{abstract}
We estimate the two-photon exchange contribution to elastic electron-proton 
scattering at large momentum transfer $Q^2$. It is shown that the leading two-photon exchange amplitude behaves as $1/Q^4$ relative to the one-photon amplitude, and can be expressed in a model independent way in terms of the leading twist nucleon distribution amplitudes. Using several models for the nucleon distribution amplitudes, we provide estimates for existing data and for  ongoing experiments. 
\end{abstract}
\maketitle 
\vspace{5cm}

Elastic electron-nucleon scattering in the one-photon ($1 \gamma$) exchange 
approximation is a time-honored tool for accessing information on the structure 
of the nucleon. Precision measurements of the proton electric to magnetic form factor ratio at larger $Q^2$ using polarization experiments~\cite{Jones00,Punjabi:2005wq,Gayou02} have revealed significant discrepancies in recent years  with unpolarized experiments using the Rosenbluth technique~\cite{Andivahis:1994rq}. As no experimental flaw in either technique has been found, two-photon ($2 \gamma$) exchange processes are the most likely culprit to explain this difference. Their study has  received a lot of attention lately, see~\cite{Carlson:2007sp} for a recent review (and references therein), and~\cite{Arrington:2007ux} for a recent global analysis of elastic electron-proton ($ep$) scattering including $2 \gamma$ corrections. 
In this work we calculate the leading in $Q^2$ behavior of the elastic $ep$ scattering amplitude with hard $2 \gamma$ exchange. 
  
\begin{figure}
\includegraphics[width = 6.75cm]{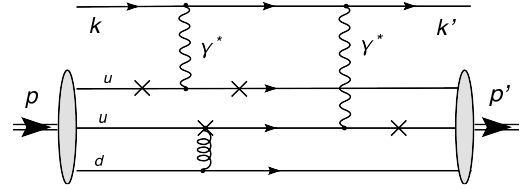}
\caption{Typical graph for the elastic $ep$ scattering with two hard photon exchanges. The crosses indicate the other possibilities to attach the gluon. The third quark is conventionally chosen as the $d-$quark. There are other diagrams where the one photon is connected with $u-$ and $d-$
quarks. We do not show these graphs for simplicity. }
\label{fig:graphs}
\end{figure}

To describe the elastic $ep$ scattering, 
$l(k)+N(p)\rightarrow l(k')+N(p')$, 
we adopt the definitions~:
$P=(p+p')/2$, $K=(k+k')/2$, $q=k-k'=p'-p$,
and choose 
$Q^{2}=-q^{2}$ and $\nu =K \cdot P$ 
as the independent kinematical invariants.  
Neglecting the electron mass, it was shown in~\cite{Guichon:2003qm} that 
the $T$-matrix for elastic $ep$ scattering can be expressed through 3 independent 
Lorentz structures as~:
\begin{eqnarray}
\label{eq:tmatrix}
T_{h, \, \lambda'_N \lambda_N} \,&=&\, 
\frac{e^{2}}{Q^{2}} \, \bar{u}(k', h)\gamma _{\mu }u(k, h)\,  \\
&&\hspace{-1.75cm} \times \, 
\bar{u}(p', \lambda'_N)\left( \tilde{G}_{M}\, \gamma ^{\mu }
-\tilde{F}_{2}\frac{P^{\mu }}{M}
+\tilde{F}_{3}\frac{\gamma \cdot K 
P^{\mu }}{M^{2}}\right) u(p, \lambda_N), \nonumber
\end{eqnarray}
where $e$ is the proton charge, $M$ is the proton mass, 
$h = \pm 1/2$ is the electron helicity and $\lambda_N$ 
($\lambda'_N$) are
the helicities of the incoming (outgoing) proton. 
In Eq.~(\ref{eq:tmatrix}), 
\( \tilde{G}_{M},\, \tilde{F}_{2},\, \tilde{F}_{3} \) are 
complex functions of \( \nu  \) and \( Q^{2} \). 
To separate the $1 \gamma$ and $2 \gamma$ exchange contributions, 
it is furthermore useful to  
introduce the decompositions~: $\tilde G_M = G_M + \delta \tilde G_M$, 
and $\tilde F_2 = F_2 + \delta \tilde F_2$, where 
$G_M (F_2)$ are the proton magnetic (Pauli) form factors (FFs) respectively, 
defined from the matrix element of the electromagnetic current, 
with $G_M(0) = \mu_p = 2.79$ the proton magnetic moment.     
The amplitudes $ \tilde{F}_{3}, \delta \tilde{G}_{M} $
and $\delta \tilde{F}_{2} $ 
originate from processes involving at least $2 \gamma$ exchange  
and are of order $e^2$ 
(relative to the factor \( e^{2} \) in Eq.~(\ref{eq:tmatrix})).
\newline
\indent
The leading perturbative QCD (pQCD) contribution to the $2 \gamma$ exchange correction to the elastic $e p$ amplitude is given by a convolution integral of
the proton distribution amplitudes (DAs) with the hard coefficient function as shown in   
Fig.~\ref{fig:graphs}. 
In the hard regime, where $ Q^{2}, s\gg M^{2}$, we calculate the amplitude in the 
Breit system, where the initial and final proton momenta correspond to two opposite 
light-like directions~:
\begin{eqnarray}
p &\simeq& Q\frac{\bar{n}}{2}, \quad \mathrm{with} \quad \bar n = (1, 0, 0, 1), 
\nonumber \\
p^{\prime} &\simeq& Q\frac{n}{2} , \quad \mathrm{with} \quad n = (1, 0, 0, -1), 
\nonumber 
\end{eqnarray}
with $(n \cdot \bar{n})=2$. The lepton kinematics are given by~:
\begin{eqnarray}
k = \zeta Q\frac{n}{2} - \bar \zeta Q \frac{\bar{n}}{2}+k_{\bot},
~k'=- \bar \zeta Q\frac{n}{2}+\zeta Q\frac{\bar{n}}{2}+k_{\bot},
\nonumber
\end{eqnarray}
where, at large $Q^2$, $\zeta$ can be determined from $s \simeq \zeta Q^2$, 
and $u \simeq - \bar \zeta Q^2$, with $\bar  \zeta \equiv 1 - \zeta$, and $\zeta \geq 1$. 
Furthermore, the transverse vector in the lepton kinematics is determined from~:
$k_{\bot}^{2}=-\zeta \bar{\zeta}Q^{2}$.

Let us now consider the proton matrix element which appears in the graph of  
Fig.~\ref{fig:graphs}. Following the notation from
\cite{Braun:2000kw}, the proton matrix element is described at leading twist level by three 
nucleon DAs as~:
\begin{eqnarray}
&&4\left\langle 0\left\vert \varepsilon^{ijk}u_{\alpha}^{i}(a_{1}\lambda
n)u_{\beta}^{j}(a_{2}\lambda n)d_{\sigma}^{k}(a_{3}\lambda n)\right\vert
p \right\rangle  \nonumber \\
&&  =V~p^{+}\left[  \left(  \frac{{\scriptsize 1}%
}{{\scriptsize 2}}\bar{n}\cdot\gamma\right)  ~C\right]  _{\alpha\beta}\left[
\gamma_{5}N^{+}\right]  _{\sigma} \nonumber \\
& & +A~p^{+}\left[  \left(  \frac{{\scriptsize 1}}{{\scriptsize 2}}\bar{n}%
\cdot\gamma\right)  \gamma_{5}C\right]  _{\alpha\beta}\left[  N^{+}\right]
_{\sigma} \nonumber \\
&&  +T~p^{+}\left[  \frac{{\scriptsize 1}}{{\scriptsize 2}}i\sigma_{\bot\bar
{n}}~C\right]  _{\alpha\beta}\left[  \gamma^{\bot}\gamma_{5}N^{+}\right]
_{\sigma},
\end{eqnarray}
with light-cone momentum $p^+ = Q$, 
where $C$ is charge conjugation matrix~: $C^{-1}\gamma_{\mu}C=-\gamma_{\mu}^{T}$, 
and where $X=\{A,V,T\}$ stand for the nucleon DAs which are defined by the light-cone 
matrix element~:
\begin{equation}
X(a_{i},\lambda p^{+})=\int d[x_{i}]~e^{-i\lambda p^{+}\left(  \sum x_{i}
a_{i}\right)  }X(x_{i}),
\end{equation}
with
\begin{equation}
d[x_{i}] \equiv dx_{1}dx_{2}dx_{3}\delta(1-\sum x_{i}).
\nonumber 
\end{equation}
In general, the following properties are valid%
\begin{eqnarray}
V(x_{1},x_{2},x_{3}) &=& V(x_{2},x_{1},x_{3}), \nonumber \\
A(x_{1},x_{2},x_{3}) &=& -A(x_{2},x_{1},x_{3}), \nonumber \\
T(x_1, x_2, x_3) &=& \frac{1}{2}\left[  V-A\right]  \left(  1,3,2\right)  +\frac{1}%
{2}\left[  V-A\right]  \left(  2,3,1\right) , 
\nonumber
\end{eqnarray}
i.e. we have only two independent functions.

In the large $Q^2$ limit, the pQCD calculation of Fig.~\ref{fig:graphs} 
involves 24 diagrams, and leads to hard $2 \gamma$ corrections to 
$\delta \tilde{G}_{M}$, and $\nu/M^2 \tilde F_3$, which are found as~:
\begin{eqnarray}
\delta \tilde{G}_{M} &=& -\frac{~~\alpha_{em}\alpha_{S}(\mu^{2})}{Q^{4}}\left(
\frac{4\pi}{3!}\right)  ^{2}  \, (2 \zeta - 1) \nonumber \\
&\times& \int d[y_{i}]~d[x_{i}]  \, \frac{4\,x_2 \, y_2}{D} 
\nonumber \\
&\times&  \left\{ {Q_u}^2 \, 
\left[ (V^\prime + A^\prime)(V + A) + 4 T^\prime T \right](3,2,1) \right. \nonumber \\
&& + Q_u Q_d \, 
\left[ (V^\prime + A^\prime)(V + A) + 4 T^\prime T \right](1,2,3) \nonumber \\
&& \left. + Q_u Q_d \, 
2 \left[ V^\prime V + A^\prime A \right](1,3,2) \right\},
\label{eq:gm}
\end{eqnarray}
and
\begin{eqnarray}
\frac{\nu}{M^2} \tilde{F}_{3} &=& -\frac{~~\alpha_{em}\alpha_{S}(\mu^{2})}{Q^{4}}\left(
\frac{4\pi}{3!}\right)  ^{2} \, (2 \zeta - 1) \nonumber \\
&\times& \int d[y_{i}]~d[x_{i}]  \, \frac{2 (x_2 \, \bar y_2 +   \bar  x_2 \,  y_2) }{D} 
\nonumber \\
&\times&  \left\{ {Q_u}^2 \, 
\left[ (V^\prime + A^\prime)(V + A) + 4 T^\prime T \right](3,2,1) \right. \nonumber \\
&& + Q_u Q_d \, 
\left[ (V^\prime + A^\prime)(V + A) + 4 T^\prime T \right](1,2,3) \nonumber \\
&& \left. + Q_u Q_d \, 
2 \left[ V^\prime V + A^\prime A \right](1,3,2) \right\},
\label{eq:f3}
\end{eqnarray}
with quark charges $Q_u = +2/3$, $Q_d = -1/3$, 
$\alpha_{em} = e^2 / (4 \pi)$, 
$\alpha_s(\mu^2)$ is the strong coupling constant evaluated at scale $\mu^2$, 
and the denominator factor $D$ is defined as~:
\begin{eqnarray}
D &\equiv& (y_1 y_2 \bar y_2) \, (x_1 x_2 \bar x_2) 
\,  \left[ x_2 \bar \zeta + y_2 \zeta - x_2 y_2 + i \varepsilon \right]  \nonumber \\
&\times& \left[ x_2 \zeta + y_2 \bar \zeta - x_2 y_2 + i \varepsilon \right].  
\end{eqnarray}
The unprimed (primed) quantities in Eqs.~(\ref{eq:gm}, \ref{eq:f3}) refer to the DAs in the initial (final) proton respectively. 
Eqs.~(\ref{eq:gm}, \ref{eq:f3}) are the central result of the present work. 
One notices that at large $Q^2$, the leading behavior for $\delta \tilde G_M$ and 
$\nu / M^2 \tilde F_3$ goes as $1/Q^4$. In contrast, the invariant 
$\delta \tilde F_2$ is suppressed 
in this limit and behaves as $1/Q^6$.  

It is interesting to point out that the scaling behavior for the $2 \gamma$ amplitude obtained in the present calculation differs from the handbag calculation of~\cite{Chen:2004tw,Afanasev:2005mp}. Whereas the present calculation gives a model independent leading behavior of 
$1/Q^4$ for the $2 \gamma$ amplitude relative to the $1 \gamma$ amplitude,  the $Q^2$ behavior of the $2 \gamma$ amplitude within the handbag calculation depends on the 
specific modeling of the generalized parton distributions.  As an example, the modified Regge parameterization considered in~\cite{Chen:2004tw,Afanasev:2005mp} leads to a calculation which is of higher twist compared to the leading pQCD calculation considered here.

To evaluate the convolution integrals in Eqs.~(\ref{eq:gm}, \ref{eq:f3}), we need to insert a 
model for the nucleon twist-3 DAs, $V$, $A$, and $T$. 
The asymptotic behavior of the DAs and their first conformal moments  were given in~\cite{Braun:2000kw} as~: 
\begin{eqnarray}
V(x_{i}) &  \simeq& 120x_{1}x_{2}x_{3} f_{N}\left[  1+r_{+}(1-3x_{3})\right]
,\nonumber \\
A(x_{i}) &  \simeq& 120x_{1}x_{2}x_{3} f_{N}~r_{-}(x_{2}-x_{1}),
\nonumber \\
T(x_{i}) &  \simeq& 120x_{1}x_{2}x_{3} f_{N}\left[  1+\frac{1}{2}\left(
r_{-}-r_{+}\right)  (1-3x_{3})\right]  ,
\end{eqnarray}
and depend on three parameters : $f_N$, $r_-$ and $r_+$. 
In this work, we will provide calculations using two models for the DAs that were discussed in the  literature. The corresponding parameters  (at $\mu=1~$GeV ) are given in Table~\ref{table:DAmodels}, and are compared with recent lattice QCD calculations (QCDSF~\cite{Gockeler:2008xv}), extrapolated to the chiral limit. One notices that the parameters $r_-$ and $r_+$ in the BLW model for the proton DA are totally compatible with the lattice results, whereas the value of the overall normalization $f_N$ for the lattice DA is about 2/3 smaller than the BLW value.  Below, we will provide calculations using the models COZ and BLW, and note that a good estimate using the lattice DA can be directly obtained from our figures by scaling the BLW result by a factor $\approx~2/3$. 

\begin{table}[ht]
\begin{center}
\begin{tabular}{|c|c|c|c|}
\hline
\hline
& $f_N $ & $r_-$   & $r_+$       \\
& ($10^{-3}$ GeV$^2$) &  &  \\
\hline \;\; COZ \cite{Chernyak:1987nu} \;\; &
$5.0 \pm 0.5$ & $ 4.0 \pm 1.5 $ & $ 1.1 \pm 0.3 $   \\
\hline \;\; BLW \cite{Braun:2006hz} \;\; & 
$5.0 \pm 0.5$ & $1.37$ & $0.35 $ \\
\hline \;\; QCDSF \cite{Gockeler:2008xv} \;\; & 
$3.23$ & $1.06$ & $0.33 $ \\
  & 
$\pm 0.06 \pm 0.09$ & $\pm 0.09 \pm 0.31$ & $\pm 0.03 \pm 0.11$ \\
\hline
\hline
\end{tabular}
\end{center}
\caption{Parameters entering the proton DA (at $\mu$ = 1~GeV) 
for two models (COZ, BLW) used in this work. For 
comparison we also show a recent lattice evaluation (QCDSF).}
\label{table:DAmodels}
\end{table}%

We next calculate the effect of hard $2 \gamma$ exchange, given through  Eqs.~(\ref{eq:gm}, \ref{eq:f3}), 
on the elastic $ep$ scattering observables. The general formulas  
for the observables including the $2 \gamma$ corrections $\delta \tilde G_M$, $\delta \tilde F_2$, and 
$\nu/M^2 \tilde F_3$ were derived in~\cite{Guichon:2003qm}, to which we refer for the corresponding expressions.  

\begin{figure}[h]
\includegraphics[width=6.5cm]{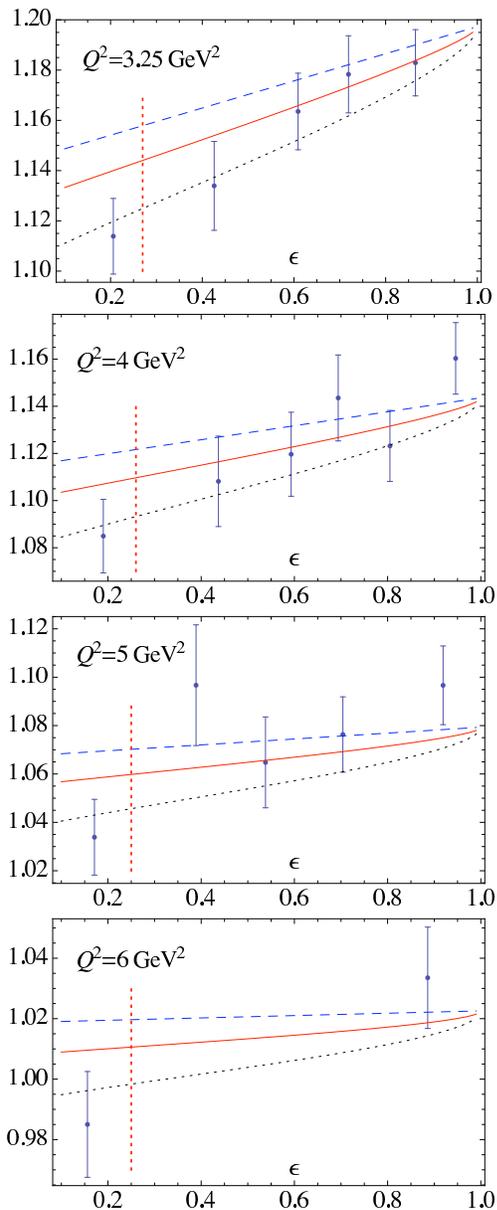}
\caption{ Rosenbluth plots for elastic $e p $ scattering: 
$\sigma_R$ divided by $\mu_p^2  / (1 + Q^2 / 0.71)^{4}$.  
Dashed (blue) curves : $1 \gamma$ exchange, using the $G_{Ep}/G_{Mp}$ ratio from polarization 
data~\cite{Jones00,Punjabi:2005wq,Gayou02}.   
Solid red (dotted black) curves show the effect including hard $2 \gamma$ exchange calculated with the BLW (COZ) model for the proton DAs. 
The vertical dotted line shows  the boundary where the $lhs$ of Eq.~(\ref{eq:appl}) is 0.5. 
The data are from Ref.~\cite{Andivahis:1994rq}.
}
\label{Fig2}
\end{figure}

In Fig.~\ref{Fig2}, we calculate the reduced cross section $\sigma_R$ as a function of 
the photon polarization parameter $\varepsilon$ and different values of $Q^2$. 
In the $1 \gamma$ exchange, $\sigma_R = G_M(Q^2) + \varepsilon / \tau~G_E(Q^2)$, with 
$\tau = Q^2 / (4 M^2)$, and the Rosenbluth plot is linear in $\varepsilon$, indicated by the dashed  straight lines in Fig.~\ref{Fig2}. The effect including the hard $2 \gamma$ exchange is shown for both the COZ and BLW models of the proton DAs. One sees that including the $2 \gamma$ exchange changes the slope of the Rosenbluth plot, and that sizeable non-linearities only occur for $\varepsilon$ close to 1. The inclusion of the hard $2 \gamma$ exchange is able to well describe the $Q^2$ dependence of the unpolarized data, when using the polarization data~\cite{Jones00,Punjabi:2005wq,Gayou02}  for the proton FF ratio $G_{Ep}/G_{Mp}$ as input. Quantitatively, the COZ model for the nucleon DA leads to a correction about twice as large as when using the BLW model. The question arises as to the applicability of the hard scattering calculation for the $Q^2$ values of the data shown in Fig.~\ref{Fig2}. On the one hand, the argument of the running coupling $\alpha_s$ is defined by the renormalization scale $\mu \sim Q$, which should be sufficiently large to validate a pQCD calculation.  On the other hand, $\tilde F_3$ contains a logarithmic singularity when $\varepsilon \to 0$ (i.e. when $\zeta \to 1$). Therefore, the applicability of the hard description is restriced by the condition~:
\begin{equation} 
\alpha_s(Q^2) \, \left| \ln (1 - 1/ \zeta) \right|  \ll 1. 
\label{eq:appl}
\end{equation}
 In Fig.\ref{Fig2} we indicate a boundary by dashed vertical lines, where 
 $\alpha_S(Q^2) \left| \ln (1- 1/\zeta) \right| \leq 0.5$, which corresponds with $\varepsilon \gtrsim 0.25$.  
 We like to note here that in contrast to the pQCD treatment of the proton FFs, which requires two hard gluon exchanges, the $2 \gamma$ correction to elastic $e p$ scattering only requires one 
 hard gluon exchange. One therefore expects the pQCD calculation to set in for $Q^2$ values in the few GeV$^2$ range, which is well confirmed by the results shown in Fig.~\ref{Fig2}. 
  
\begin{figure}[h]
\includegraphics[width=7.25cm]{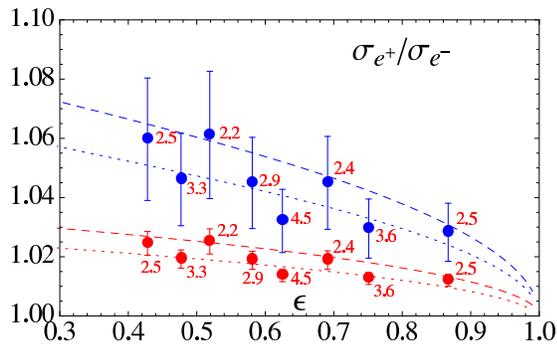}
\caption{Predictions for the ratio $\sigma_{e^{+}p}/\sigma_{e^{-}p}$ for
different values of $Q^{2}$ (shown by numbers in GeV$^{2}$) and $\varepsilon$ planned by the Olympus@DESY~\cite{olympus} experiment (we only show kinematics for which $Q^2 > 2$~GeV$^2$). 
The upper blue (lower red) points correspond with the COZ (BLW) models. 
The error-bars show the theoretical uncertainties due to the parameters
of the proton DAs and the running coupling scale, as described in the text. 
For comparision, we also show the theoretical curves for the average 
values $Q^{2}=2.4$~GeV$^{2}$ (dashed curves) and $Q^{2}=3.25~$GeV$^{2}$ (dotted curves) for both the COZ (two upper curves) and BLW (two lower curves) models. 
}%
\label{Fig3}
\end{figure}

The real part of the $2 \gamma$ exchange amplitude can be accessed directly 
as the deviation from unity of the ratio of $e^+ / e^-$ 
elastic scattering. The precision of past experiments performed at SLAC~\cite{Mar:1968qd}, was not sufficient to see a clear deviation from unity over a large range in $\varepsilon$.  
Presently, several new experiments are planned or are underway at  
VEPP-3~\cite{Arrington:2004hk}, JLab/CLAS~\cite{clas}, and 
Olympus@DESY~\cite{olympus} to make precision measurements of the $e^+/e^-$ ratio in 
elastic scattering off a proton. 
In Fig.~\ref{Fig3}, we show the predictions for the 
$\sigma_{e^{+}p}/\sigma_{e^{-}p}$ ratio for
different values of $Q^{2}$ and $\varepsilon$ 
planned by the Olympus@DESY experiment~\cite{olympus}. In order to make a comparison with our pQCD calculations, we only show kinematics for which $Q^2 > 2$~GeV$^2$.  
We also made an estimate of the theoretical uncertainty of our predictions. For this purpose, we varied  the normalization scale $\mu$ (entering $\alpha_s$) and the normalization $f_N$ (in units $10^{-3}$~GeV$^2$) for both COZ and BLW models of the proton DAs over the ranges~: $Q^2 /2 < \mu^2 < Q^2$, and $4.5 < f_N  < 5.5$. Furthermore, for the COZ model, we also varied the parameters $r_+$ and $r_-$ in the range~: 
$0.8 < r_+ < 1.4$, and $2.5 < r_- < 5.5$. The resulting theoretical error estimate is also indicated on 
Fig.~\ref{Fig3}. 
The Olympus@DESY experiment aims at a statistical precision of the $e^+ / e^-$ cross section ratio of better than one percent for an average $Q^2 = 2.2$~GeV$^2$.  Our calculations predict a deviation from unity for these ratios in the range 2.5\% (BLW) to 5\% (COZ), allowing for an unambiguous test with the upcoming measurements.    

\begin{figure}
\includegraphics[width=7.25cm]{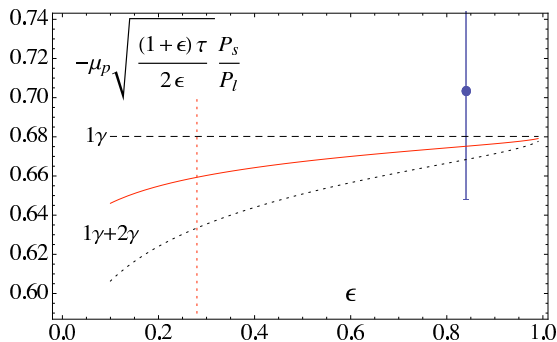}
\caption{ The ratio $P_{s}/P_{l}$ as a function of $\varepsilon$ for
$Q^{2}=2.5$~GeV$^{2}$. The  horizontal curve is the result of the $1 \gamma$ calculation. The dotted black (solid red) curves correspond to the $1 \gamma + 2 \gamma$ calculations using the COZ (BLW) models for the proton DAs. 
The data point is from the JLab/Hall A experiment~\cite{Jones00,Punjabi:2005wq}. 
}
\label{Fig4}
\end{figure}

In Fig.~\ref{Fig4}, we show the ratio of the proton recoil polarization components $P_s / P_l$ as a function of $\varepsilon$ for $Q^2 = 2.5$~GeV$^2$. The $1 \gamma$ prediction is given by the horizontal line. Our $1 \gamma + 2 \gamma$ prediction yields a negative correction which increases with decreasing $\varepsilon$. Around $\varepsilon = 0.3$ it reduces the $P_s / P_l$ ratio by 2 \% 
for the BLW model, and by  around 4 \% for the COZ model. The JLab/Hall A experiment~\cite{Jones00, Punjabi:2005wq} has measured this ratio at a large $\varepsilon$ value around 0.85. A new JLab/Hall C experiment~\cite{hallc}, which is currently under analysis, has recently measured this ratio 
at $Q^2 = 2.5$~GeV$^2$ for three $\varepsilon$ values between $0.15$ and $0.8$.  The expected experimental precision of around 1\% for this ratio, will therefore allow to test our predictions, which are in the $2 - 5$~\% range.  

The observables discussed above test the real parts of the $2 \gamma$ amplitudes.  
The imaginary parts of Eqs.~(\ref{eq:gm}, \ref{eq:f3}), arising from lepton propagator singularities, can be tested by polarizing the target or recoiling proton perpendicular to the scattering plane. 

In summary, we calculated the leading in $Q^2$ behavior of the 
$2 \gamma$ exchange contribution to elastic $ep$ scattering. It was found 
that the leading $2 \gamma$ amplitude is given by processes involving one hard gluon exchange, resulting in a  $1/Q^4$ behavior of the $2 \gamma$ amplitude relative to the 
$1 \gamma$ amplitude. We expressed the leading $2 \gamma$ amplitude in terms of the leading twist nucleon DAs.  Using two models for the nucleon DAs, we found that,  for $Q^2$ in the few GeV$^2$ range,  these calculations can quantitatively explain the 
slope of the Rosenbluth plot when using the $G_{Ep}/G_{Mp}$ polarization data as input. Furthermore, we have shown that ongoing and planned elastic $e p$ scattering experiments both for the $\varepsilon$ dependence of the recoil polarization ratio $P_s/P_l$ as well as for the $e^+/e^-$ ratio, have the precision to test our predictions.

We like to thank M. Polyakov for discussions. 
This work was supported by the BMBF, by the german DFG, and 
the U.S. DOE under contract DE-FG02-04ER41302.

\end{document}